\documentclass[%
  preprint,
 aip,
 amsmath,amssymb,
  jcp,
]{revtex4}

\usepackage{graphicx}
\usepackage{dcolumn}
\usepackage{bm}

\begin{document}


\title{Role of coherence in the plasmonic control of molecular absorption}

\author{Emanuele Coccia}
\altaffiliation{Present address: Department of Chemical and Pharmaceutical Sciences, University of Trieste, via Giorgieri 1, Trieste, Italy}
\affiliation{Department of Chemical Sciences, University of Padova, via Marzolo 1, Padova, Italy}
  
\author{Stefano Corni}%
 \email{stefano.corni@unipd.it}
\affiliation{Department of Chemical Sciences, University of Padova, via Marzolo 1, Padova, Italy}%
\affiliation{CNR Institute of Nanoscience, via Campi 213/A, Modena, Italy}

\date{\today}

\begin{abstract}

The interpretation of  nanoplasmonic effects on molecular properties, such as metal-enhanced absorption or fluorescence, typically assumes a fully coherent picture (in the quantum-mechanical sense) of the phenomena. Yet, there may be conditions where the coherent picture breaks down, and decoherence effect should be accounted for. Using a state-of-the-art multiscale model approach able to include environment-induced dephasing, here we show that metal nanoparticle effects on the light absorption by a nearby molecule is strongly affected (even qualitatively, i.e., suppression vs enhancement) by molecular electronic decoherence. The present work shows that decoherence can be thought as a further design element of molecular nanoplasmonic systems.


\end{abstract}

\maketitle


\section{Introduction}

 The electronic and optical properties of a molecule interacting with an external electromagnetic field are strongly modified by the presence of  metal nanoparticles (NPs) \cite{ger81, van04,bk:np,cor13,wil16}. Among the striking phenomena induced by plasmonic effects, modulation (enhancement or quenching) of molecular fluorescence and absorption has been extensively investigated from both experimental and theoretical sides, also for systems as complex as light-harvesting proteins \cite{dul02,dul05,kha14,pia16,and04,ang06,nl072854o,nl100254j,nie10,nl202772h,ang13,and13,Wientjes2014,Wientjes2014b,and15,Cohen:2015bd,cap16}. \\
It is well known that the optical response of the molecule depends on several aspects, determining the global plasmonic-induced enhancement or quenching: the type and polarization of the applied electromagnetic field, the nature, size and shape of the NP \cite{and04,vuk09}, the mutual orientation and distance between molecule and NP \cite{car06,vuk09}, and effects from the surrounding environment. 
A point that was not explored in the past is the role of dephasing, which determines loss of coherence in the quantum state of the system \cite{bau14,str17}. 
Understanding and controlling the role of coherence in ultrafast molecular processes \cite{sch17}, including biological systems \cite{lee07,eng07,ish09,col10,sch10}, materials \cite{sun15,mul18} and also in quantum computing \cite{lad10}, represents nowadays a key aspect at the forefront of physical chemistry research. 

In this work, we apply a multiscale time-resolved approach to unravel the interplay occurring between plasmonic effects and electronic coherence, in tuning metal-affected molecular absorption.
\noindent In order to accomplish that, the time-dependent stochastic Schr\"odinger equation \cite{bk:open,sse1,dal14,open,bk:kam,ec18} (SSE) approach has been applied to simulate the dephasing of the electronic molecular state induced by a surrounding environment, and to include in the model spontaneous (NP-modified) emission and nonradiative decay. 
We have used LiCN as model chromohore, described at configuration interaction with singly-excited configurations (CIS) level of theory, and a spherical silver NP with a radius of 5 nm. More complex molecules, such as dyes,\cite{ec18} or NPs of different shapes\cite{pip16} are accessible with our approach. Yet, the simplicity of this setup allowed us to identify the best condition to highlight the role of coherence. Three main ingredients are employed here:
\begin{itemize}
\item
a realistic description of the molecule, using standard quantum chemistry techniques; 
\item
 a classical representation of NP using an electromagnetic description. The mutual polarization between molecule and NP is simulated with a time-dependent boundary element approach \cite{pip16}: NP is polarized by the external electromagnetic field and by the field generated by the time-evolution of the electronic density of the molecule. The interaction between the molecule and NP is purely electromagnetic. Polarization of the NP surface is simulated by a surface mesh \cite{pip16}. The multiscale model used to describe the interaction between the molecule and the NP has been extensively explained elsewhere \cite{pip16}, and takes into account the local field spatial dependence.\cite{Neuman2018} Various theoretical and computational models have been proposed through the years to study molecular plasmonics, especially in the frequency domain, exploiting  a multiscale representation of the molecule+NP system\cite{jen08,mor11,men19}; 
\item Markovian SSE to get the time evolution of the density matrix of LiCN, perturbed by the presence of  NP; SSE is propagated using a quantum jump algorithm \cite{qjump1,qjump2,qjump3,revqjump}.
\end{itemize}

In Section \ref{the} we have briefly reviewed the theoretical and computational strategy, based on SSE and classical electromagnetic description of NP; computational details are also given. Results are presented and discussed in Section \ref{res}, while conclusions are reported in the last Section.

\begin{figure}[tbp]
\includegraphics[width=0.45\textwidth]{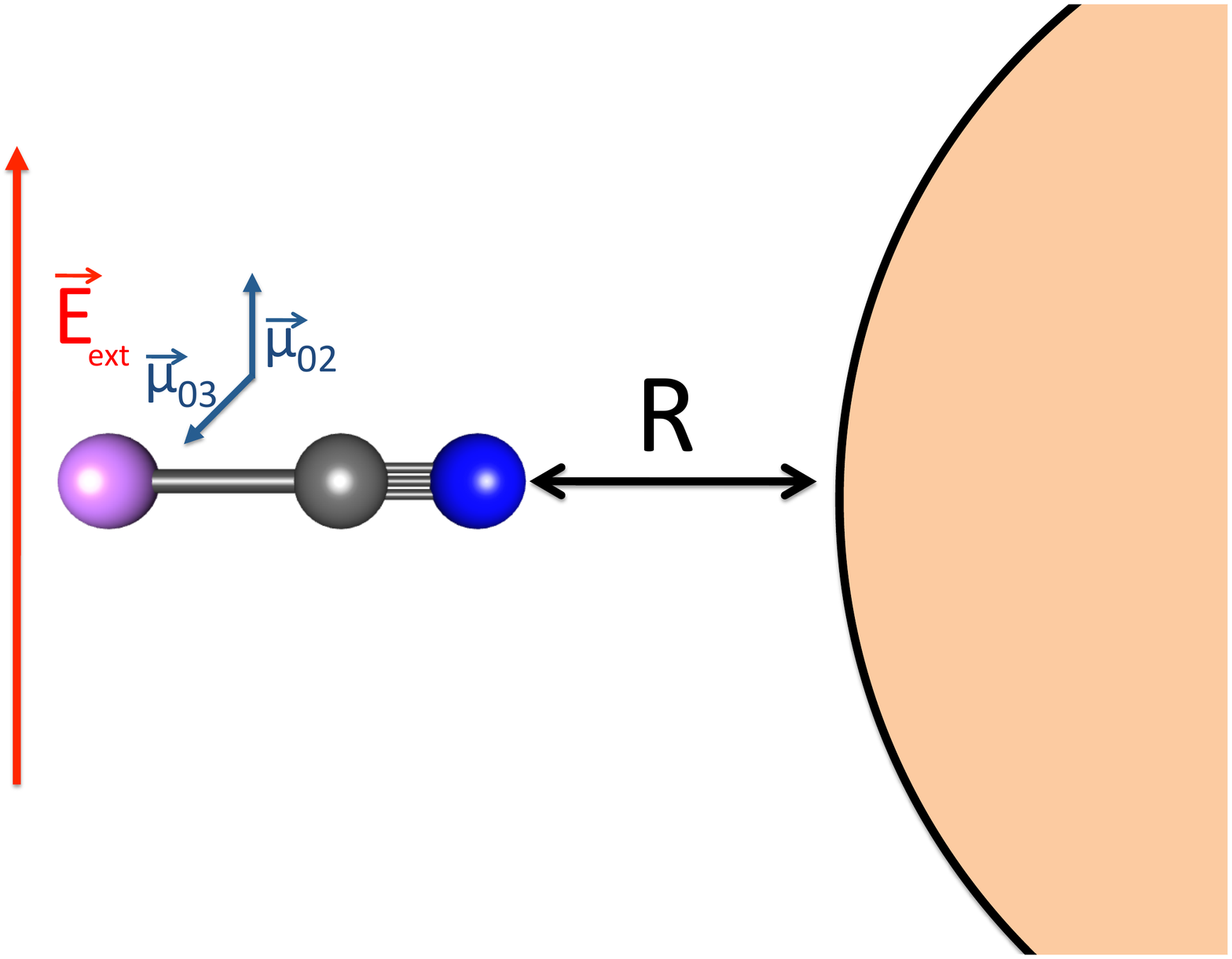}
\caption{Schematic representation of the LiCN+NP system. The external field $\vec{E}_{ext}$ is perpendicular to the molecular axis. R is the distance between the nitrogen atom and the NP surface. Transition dipole moments $\vec{\mu}_{02}$ and $\vec{\mu}_{03}$ of LiCN are also reported. Drawing out of scale.\label{fig1} }
\end{figure}

\section{Theory and computational details}
\label{the}

\subsection{Stochastic Schr\"odinger equation}

The stochastic Schr\"odinger equation (SSE) \cite{sse1} in the Markovian limit (employed in the present work) is given by \cite{sse1}:
\begin{equation}
i \frac{d}{dt} |\Psi_{S}(t) \rangle = \hat{H}_{S}(t) |\Psi_{S}(t) \rangle +  \sum_{q}^M l_{q}(t) \hat{S}_{q} |\Psi_{S}(t) \rangle -  \frac{i}{2} \sum_{q}^M \hat{S}_{q}^{\dagger}\hat{S}_{q} |\Psi_{S}(t) \rangle,
\label{eq:sse}
\end{equation}
where $|\Psi_{S}(t) \rangle$ is the system (i.e., LiCN electronic structure) wave function, $\hat{H}_{S}(t)$ is the time-dependent system Hamiltonian (defined below in Eq. \ref{eqh}), $\hat{S}_{q}$
describes the effect of the external bath on the system, through the $M$ interaction channels $q$.\\
The non-Hermitian term $- \frac{i}{2} \sum_{q}^M \hat{S}_{q}^{\dagger}\hat{S}_{q}$ is the dissipation due to the environment, whereas $\sum_{q}^M l_{q}(t) \hat{S}_{q}$ is the fluctuation term, modeled by a Wiener process $l_{q}(t)$, i.e. a white noise associated to the Markov approximation. \\
Diagonal and off-diagonal elements of the reduced density matrix $\hat{\rho}_S(t)$, respectively populations and coherences of the states of the system at time $t$, are obtained  by averaging on the number of independent realizations $N_{\text{traj}}$ of SSE \cite{ec18}. Given the definition of the reduced density matrix 
\begin{equation}
\hat{\rho}_S(t) \equiv \frac{1}{N_{\text{traj}}} \sum_j^{N_{\text{traj}}} \vert \Psi_{S,j}(t) \rangle \langle \Psi_{S,j}(t) \vert,
\end{equation}
where $|\Psi_{S,j}(t) \rangle$ is the system wave function corresponding to $j$-th realization, and expanding $\vert \Psi_{S,j}(t) \rangle$ into stationary eigenstates $| m \rangle$ of the system
\begin{equation}
|\Psi_{S,j}(t) \rangle = \sum_m^{N_\text{states}} C_{m,j}(t) | m \rangle,
\label{eq:cis}
\end{equation}
one defines population and coherences as the following:
\begin{widetext}

\begin{eqnarray}
\text{population of state} \quad q & \equiv & (\hat{\rho}_S(t))_{qq} = \frac{1}{N_{\text{traj}}} \sum_j^{N_{\text{traj}}} \vert C_{q,j}(t) \vert^2 \\
\text{coherence of states} \quad q \quad \text{and} \quad k & \equiv& (\hat{\rho}_S(t))_{qk} = \frac{1}{N_{\text{traj}}} \sum_j^{N_{\text{traj}}} C_{q,j}^*(t) C_{k,j}(t). 
\end{eqnarray}

\end{widetext}

For a large number of trajectories $N_{\text{traj}}$, SSE outcomes (in the Markov limit) coincide with the results from Lindblad equation \cite{ec18}.

\subsection{Classical electrodynamics for NP}

The time-dependent Hamiltonian $\hat{H}_s(t)$ of the system molecule+NP is defined as
\begin{equation}
\hat{H}_S(t) = \hat{H}_0 - \vec{\hat{\mu}} \cdot \vec{E}_{ext}(t) + (\mathbf{q}_{ref}(t) + \mathbf{q}_{pol}(t)) \cdot \hat{\mathbf{V}},
\label{eqh}
\end{equation}
where $\hat{H}_0$ is the time-independent electronic Hamiltonian, $\vec{\hat{\mu}}$ is the molecular dipole, $\mathbf{q}_{ref}(t)$ and $\mathbf{q}_{pol}(t)$ are the charges on the NP surface determined by the direct polarization due to the external field ($\mathbf{q}_{ref}(t)$) and to the time evolution of the molecular electronic density ($\mathbf{q}_{pol}(t)$), and $\hat{\mathbf{V}}$ is the molecular electrostatic potential calculated at the positions of the charges \cite{pip16}. \\
A time-dependent version of the boundary element method has been used to describe the time-evolution of the interaction between the molecular electrostatic potential and apparent charges located on the NP surface \cite{pip16}. \\ 
A Drude dielectric function has been used in this work:
\begin{equation}
\epsilon(\omega) = 1 - \frac{\Omega_{p}^{2}}{\omega^{2} + i \gamma \omega}
\end{equation}
with $\Omega_{p}$  and $\gamma$ the plasma frequency of the metal and the relaxation time, respectively \cite{pip16}.

\subsection{CIS expansion of the wave function}

 SSE is coupled to a quantum-chemistry description of the molecular target. In the expansion of Eq. \ref{eq:cis} 
$C_m(t)$ are time-dependent expansion coefficients, and $| m \rangle$ represents the $m$-th time-independent CIS eigenstate of the isolated system, with eigenvalue $E_m$. \\ 
By defining
$\hat{H}_{SSE}(t) \equiv \hat{H}_{S}(t) + \sum_{q}^M l_{q}(t) \hat{S}_{q} -  \frac{i}{2} \sum_{q}^M \hat{S}_{q}^{\dagger}\hat{S}_{q} $, the matrix form of the SSE is formally given by
\begin{equation}
i\frac {\partial \textbf{C}(t)} {\partial t} = \textbf{H}_{SSE}(t) \textbf{C}(t), 
\end{equation}
where $\textbf{C}(t)$ is the vector of the time-dependent expansion coefficients and $\textbf{H}_{SSE}(t)$ is the matrix representation at time $t$ of $\hat{H}_{SSE}(t)$ in the basis of the CIS eigenstates ($\textbf{H}_{SSE}(t))_{kl} = \langle k | \hat{H}_{SSE}(t) | l \rangle$. Coefficients $\textbf{C}(t)$ are propagate via a second-order Euler algorithm \cite{ec18}.

\subsection{$\hat{S}_q$ operators}

Interaction channels used in our simulations imply relaxation (spontaneous emission and nonradiative decay) and pure electronic dephasing. \\
Relaxation refers to the decay from an electronic excited state of the fluorophore to its ground state $\vert 0 \rangle$ \cite{tre08}. The following operator has been used:
\begin{equation}
 \hat{S}_q^{\text{rel}} = \sqrt{\Gamma_q} | 0 \rangle \langle q |.
 \label{eq:rel}
\end{equation} 
As a result, the population $|C_q(t)|^2$ of each state $q$ in the given trajectory exponentially decays with a rate $\Gamma_q$. $\Gamma_q$ for spontaneous emission is proportional to the square of the total transition dipole moment, i.e the sum between the transition dipole moment of LiCN interacting with NP, and the dipole induced in NP \cite{and04}. For the nonradiative decay, $\Gamma_q$ value is given in the model as external parameter.  \\
Dephasing acts on the decay of the off-diagonal elements of the reduced density matrix, i.e. the coherences of the system. 
We have used here an operator for the pure dephasing that changes the sign of the element $|q \rangle \langle q | $ \cite{ec18}:
\begin{equation}
\hat{S}_{q}^{\text{dep}} = \sqrt{\gamma_{q}/2} \sum_p^{N_\text{states}}  M(p,q) | p \rangle \langle p |,
\label{eq:dep}
\end{equation}
where $M(p,q)$ is equal to -1 if $p = q$ or equal to 1 otherwise. As explained in Ref. \cite{ec18}, this operator guarantees the population of the various states remains unchanged during the propagation \cite{ec18}. Values of $\gamma_q$ are input parameters.

\subsection{Computational details}

CIS calculations on LiCN equilibrated with NP for energy eigenvalues, electrostatic potential of the molecule and transition dipole moments have been carried out using a locally modified version of Gamess \cite{gam1,gam2}. A 6-31G(d) basis set has been employed \cite{tre08,pip16}. The geometry of the LiCN molecule used in the calculations is the one of Refs \cite{tre08,pip16} (Li-C and
C-N bond lengths of 1.949 and 1.147 \AA, respectively). 
The following parameters have been chosen for the silver NP: plasma frequency $\Omega_p$ = 0.332 au = 9.03 eV 
and $\gamma = 1.515 \times 10^{-3} \text{au} = 4.123 \times 10^{-2}$ eV. Plasmon excitation energy is equal to 5.28 eV. 370 tesserae have been used for the boundary element method. \\
The first 10 excited states are kept in the expansion of the time-dependent wave function. A nonradiative decay time of 1 ps and various dephasing times (5 fs, 50 fs and 1 ps) have been chosen in our simulations.  \\
For all the distances between LiCN and NP, and for the isolated molecule, dynamics of 1 ps have been carried out.
The $\delta$ pulse is simulated with a Gaussian centered at 157 as with an amplitude of 48 as; the intensity is 0.1 W/cm$^2$. A time step of 1.21 as has been employed in all the simulations.
The $\delta$ pulse has been simulated by a Gaussian function with a FWHM of 80 as.
Calculations have been also carried out using a pulse with a Gaussian envelope, resonant with the excitation of interest of the bare LiCN molecule
 \begin{equation}
\vec{\text{E}}_{\text{ext}}(t)= \vec{\text{E}}_{\text{max}} \exp \left   (  - \frac{(t-t_{0})^{2}}{2\sigma^{2}}   \right ) \sin(\omega t),  
\end{equation}
where $\vec{\text{E}}_{\text{max}}$ is the maximum field amplitude, $t_0$ is the center of the Gaussian function, $\sigma$ is the Gaussian amplitude corresponding to FWHM = 30 fs, and $\omega = 6.580$ eV is the pulse frequency, resonant with the excitation energy to the second (and third) excited state, used as a reporter of the molecular absorption. Evolution of ground-state and excitation ($\vert 0 \rangle \rightarrow \vert 2(3) \rangle $) energy with the N distance is reported in Figure 1 and 2 of Supporting Information (SI), respectively. \\
The real-time propagation of the wave function  using SSE has been performed using the homemade WaveT code \cite{pip16}.




\begin{figure}[tbp]
\includegraphics[width=0.48\textwidth]{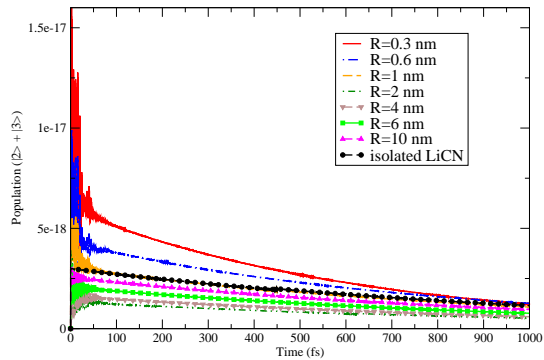}
\caption{Time evolution of the sum of the populations of the two degenerate $\vert2\rangle$ and $\vert3\rangle$ states for different distances R between the nitrogen atom of LiCN and the NP surface. The order of magnitude of the statistical error (not shown) is approximately 10$^{-19}$. Nonradiative decay time equal to 1 ps and pure dephasing time equal to 50 fs. \label{fig2} }
\end{figure}

\section{Results and discussion}
\label{res}

A time-resolved approach permits to analyze the time evolution of the properties of interest, as the population of the excited states, and how they interact with the external field and with physical processes (plasmon decay, electronic dephasing, nonradiative decay etc.) occurring at different time scales. \\
We first have applied a broadband $\delta$ pulse on both molecule and NP, with the field polarization being perpendicular to the molecular axis, which is in turn perpendicular to the NP surface (Figure \ref{fig1}). Nitrogen is the closest atom to the NP surface. Our results have been obtained in weak coupling and linear (field intensity of 0.1 W/cm$^2$) regime. The applied field strongly interacts with the molecule due to the high transition dipole moments of the degenerate second and third excited states, $|2\rangle$ and $|3\rangle$, of LiCN, as shown in Figure \ref{fig1}. We therefore focus on the time evolution of the sum of the populations of the $|2\rangle$ and $|3\rangle$ states. \\
We have started by computing the time evolution of the molecular populations along an electron dynamics of 1 ps,  using a nonradiative relaxation time of 1 ps and a pure dephasing time $T_2$ equal to 50 fs (a reasonable value for molecules)\cite{hil11}, over a range of LiCN-NP distance R between 0.3 and 10 nm (Figure \ref{fig2}). Radiative relaxation decay due to metal-enhanced spontaneous emission is also included via the approach detailed in Refs\cite{and04,vuk09}, although it occurs on a much longer time scale (i.e., nanoseconds) than that we are focusing on. 

Two distinct features are clear from the results in Figure \ref{fig2}: a rapid variation of the populations is seen at short times ($<$ 100 fs), while the decay at longer times is dominated by the nonradiative decay of 1 ps. The complex behaviour for times $<$ 100 fs is due to the interaction between molecular and plasmon excitations. The external electric field pulse is in fact exciting both the molecule and the plasmon, and the electromagnetic field produced by the latter is felt by the molecule till it decays by Landau damping. We also notice that, for the LiCN-NP distances R= 0.3 nm and 0.6 nm, the LiCN population is larger at any time than that obtained for the isolated molecule. On the contrary, for larger distances the populations are smaller than those achieved for the isolated molecule (after the initial 100 fs transient regime).
To further characterize this observation, we focus on the trend of the populations as a function of the distance for a specific time after the pulse. In particular, we have chosen to analyze populations at 500 fs, a time long enough for the NP-induced ultrafast processes to have already occurred.   

\begin{figure}[tbp]
\includegraphics[width=0.43\textwidth]{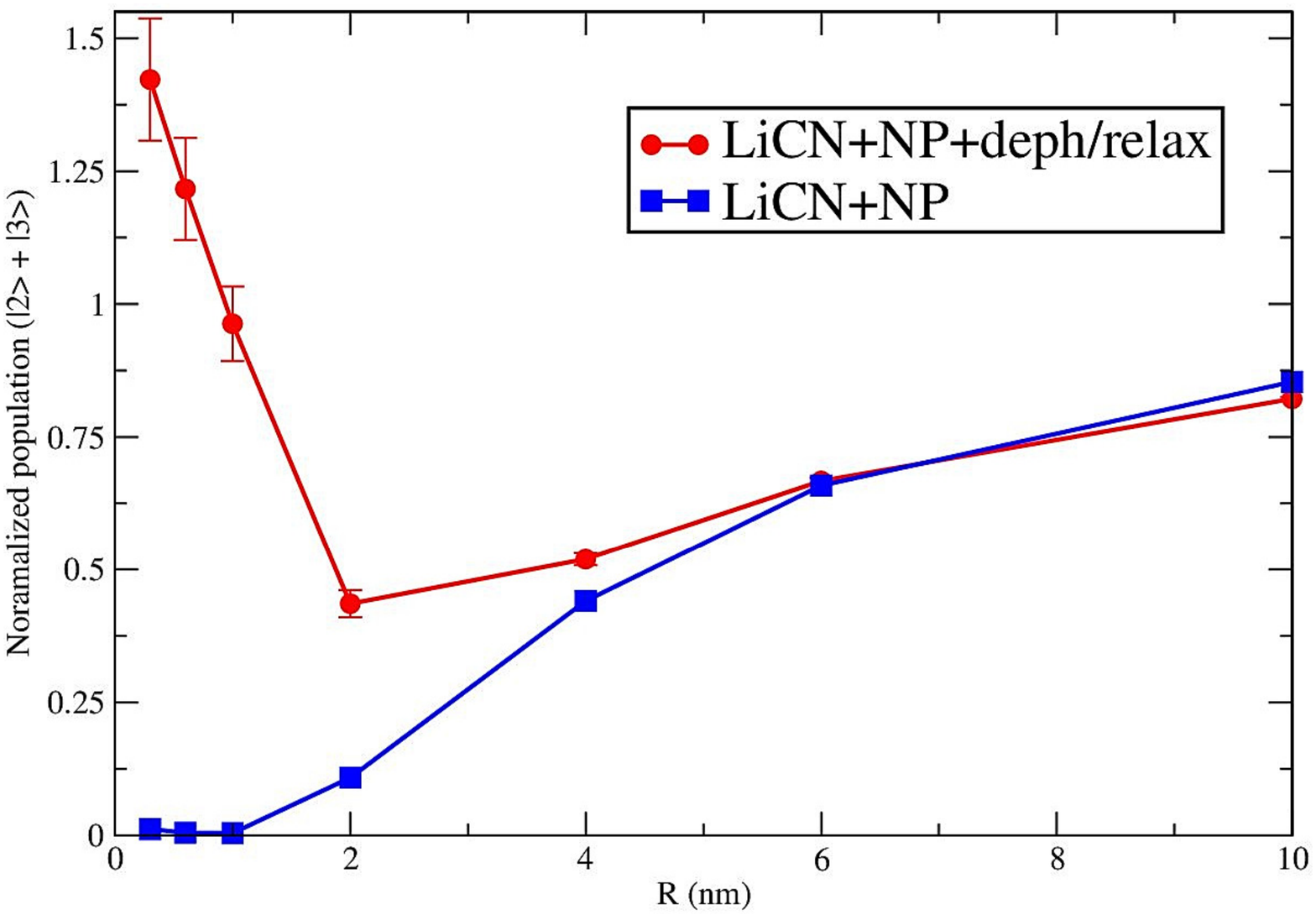}
\caption{Sum of the normalized (see text) population of the $\vert2\rangle$ and $\vert3\rangle$ states at 500 fs as a function of the distance R. Nonradiative decay time equal to 1 ps and dephasing time $T_{2}$ of 50 fs.\label{fig3}}
\end{figure}

The value of the population as a function of the distance R at 500 fs, extracted from Figure \ref{fig2} is reported in Figure \ref{fig3} (red curve "LiCN+NP+deph/relax"). The non-monotonic trend noticed above is here evident. To evaluate the role of the decoherence accounted for such results, we have to compare them to what can be obtained without decoherence (pure coherent condition, "LiCN+NP" blue curve in Figure \ref{fig3}).
Population values are normalized with respect to the respective values without NP. The two curves have significantly different trends. This is the most relevant finding of the present work: electronic decoherence (with a reasonable $T_2$ of 50 fs) can dramatically alter the plasmon effects on the absorption of the molecule, to a level that should be amenable of experimental detection. We have now to clarify where this large difference comes from. The first step in the explanation is to understand the peculiar feature of the coherent result trend. The latter presents a strong quenching of the absorption at small distances, with a recovery as the distance increases. The origin of this trend is akin to the phase induced radiative rate suppression,\cite{and04,dul05} but referred to absorption: The external field $\vec{E}_{\text{ext}}$ (a $\delta$ pulse) excites the molecule with a given phase, and also controls the excitation phase of the plasmon. The field produced by the plasmon, $\vec{E}_{\text{ind}}$, is out of phase with respect to that acting on the molecule for this specific molecule-NP setup and plasmon frequency. As such, its global effect is to take back the excited state population to the ground state. As long as the molecular electronic dynamics is coherent, the fact that the initial pulse and the plasmon oscillations happen on different time intervals is immaterial, like in two pulse phase-locked ultrafast spectroscopies.\cite{hil11} Of course the  intensity of such unfavorable plasmon electric field acting locally on the molecule will decrease by increasing the molecule-NP distance, which explains the population recovery at large distances. 
To put this discussion on a more formal ground, we note that the coherent regime corresponds to the condition employed by Gersten and Nitzan \cite{ger81} to derive the theoretical model for the enhancement or quenching of the fluorescence of a single molecule interacting with a NP. Defining $\gamma_{exc}$ and $\gamma_{exc}^0$ as the excitation rate of LiCN in presence of NP and for the bare molecule, the normalized excitation rate $\gamma_{exc}^{coh}$ is given for the coherent regime by \cite{ger81,ang06}
\begin{equation}
\gamma_{exc}^{coh} \equiv
\frac{\gamma_{exc}}{\gamma_{exc}^0} = \frac {\vert \vec{n}_{\mu} \cdot ( \vec{E}_{ext} + \vec{E}_{ind} )\vert^2}  { \vert \vec{n}_{\mu} \cdot \vec{E}_{ext} \vert^2},
 \label{eq:coh}
\end{equation}
where $\vec{n}_{\mu}$ is the unit vector pointing in direction of the transition dipole moment $\vec{\mu}$. The sum appearing inside the square modulus in Eq.\ref{eq:coh} is where coherence of the two fields is used. The fields can cancel each other, like in the numerical example discussed so far.

When a finite decoherence time is accounted for, the fact that the pulse and the plasmon-induced electric field act on different time scales does matter. In particular, to explain the enhanced absorption at short distances when a dephasing time $T_2$ of 50 fs is included in the calculation ("deph/relax" curve in Figure \ref{fig3}), one has to consider the interplay between dephasing, plasmon and molecular excitation. In fact, decoherence of the electronic state of the molecule "decouples" the contributions from $\vec{E}_{\text{ext}}$ and $\vec{E}_{\text{ind}}$, which sum up incoherently regardless of the value of the relative phase. Therefore, now the effects of the two fields do not cancel each other anymore, and the resulting excitation rates will sum up to give the normalized excitation rate $\gamma_{exc}^{inc}$ in presence of electronic dephasing

\begin{equation}
\gamma_{exc}^{inc} \equiv
  \frac {\vert \vec{n}_{\mu} \cdot  \vec{E}_{ext} \vert^2 + \vert \vec{n}_{\mu} \cdot  \vec{E}_{ind} \vert^2}  { \vert \vec{n}_{\mu} \cdot \vec{E}_{ext} \vert^2}.
  \label{eq:inc}
\end{equation}
Clearly $\gamma_{exc}^{inc}$ is larger than one, i.e., absorption suppression is not possible, only absorption enhancement can take place. This is the extreme case of fully incoherent action (we shall see the numerical results later in this work). The case in red in  Figure \ref{fig3} still shows  a partial coherence. As a result, a dip in the absorption rate due to phase-induced suppression is still present, only  the minimum is pushed at longer distances and the suppression is lower.

\begin{figure}[tbp]
\includegraphics[width=0.56\textwidth]{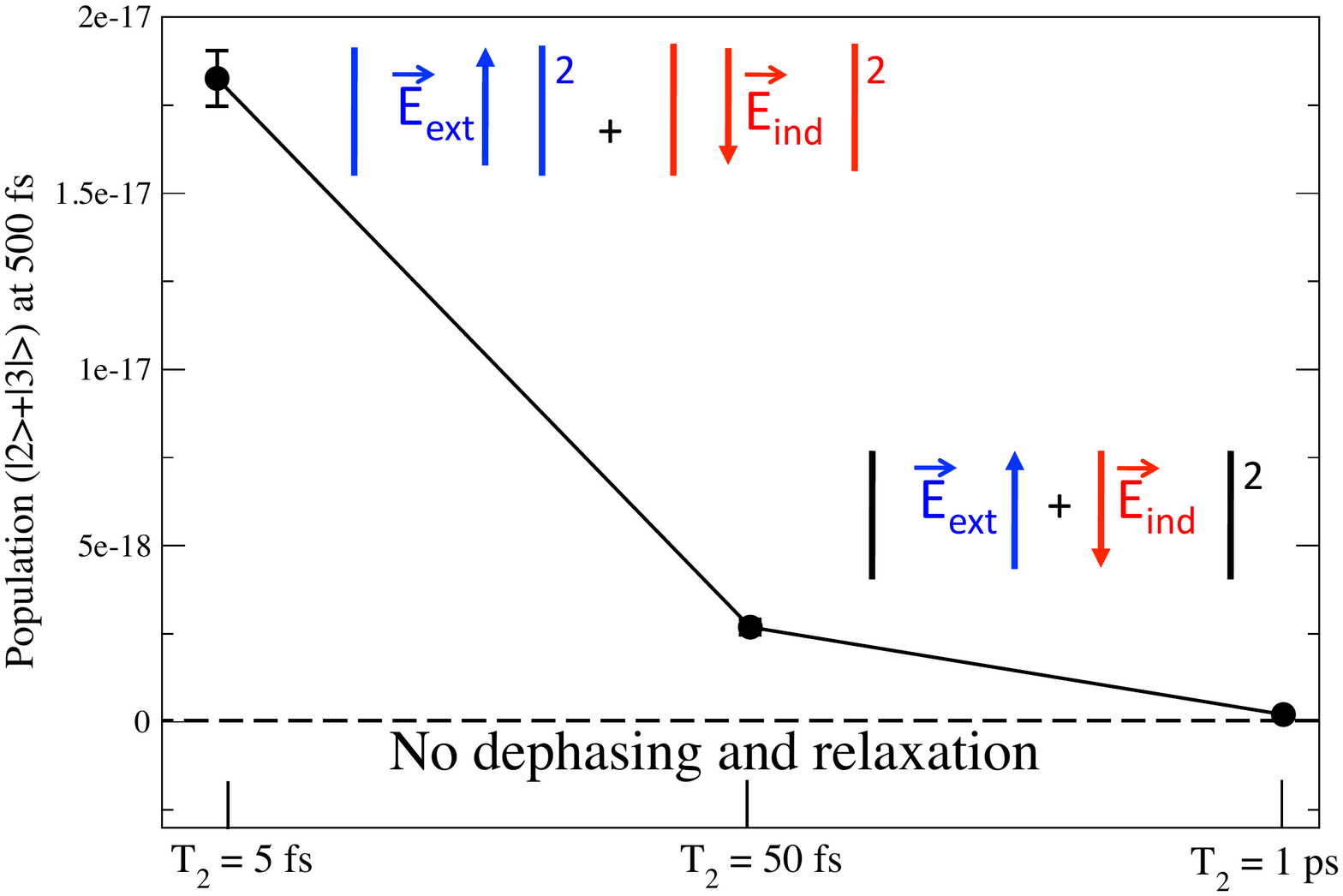}
\caption{Sum of the population of the $\vert2\rangle$ and $\vert3\rangle$ states at 500 fs for R=0.3 nm and with $T_2$ = 5 fs, 50 fs and 1 ps. \label{fig4} }
\end{figure}

Once we have established that decoherence contributes to tune the molecular absorption, the next step is to have a quantitative insight into this dephasing- and plasmon-mediated process. 
Plasmon decay times $T_{plas}$ for this type of NP typically span the range 10-100 fs; in our simulations $T_{plas}\approx50$ fs. Qualitatively, the effect of the plasmon excitation is observed in the first 100 fs of the electron dynamics (Figure \ref{fig2}), where strong rapid oscillations are due to a beat between the molecular excitation and the plasmon. \\
Focusing on the LiCN absorption at the shortest distance (R=0.3 nm) with respect to NP, we have investigated the role of the dephasing regime on the molecular excitation by computing the sum of the $\vert 2 \rangle$ and $\vert 3 \rangle$ populations when changing the dephasing time $T_2$. Three dephasing conditions have been specifically explored (Figure \ref{fig4}): a fast dephasing, $T_2=5$ fs $< T_{plas}$, a slow dephasing, $T_2=1$ ps $>> T_{plas}$, and an intermediate regime, $T_2 =$ 50 fs $\sim T_{plas}$ (used for the calculations reported in Figure \ref{fig3}). 
The largest population (taken at 500 fs) is found when the fastest dephasing ($T_2=5$ fs) is applied, while for the slowest dephasing ($T_2 = 1$ ps) one retrieves the value obtained for the fully coherent LiCN+NP system. Time evolution of the  ($\vert 2 \rangle$ + $\vert 3 \rangle$) populations for the different coherence regimes are collected in Figure 3 of SI.
This is fully in line with the explanations given in the previous paragraph: depending on $T_2$ and $T_{plas}$, external and plasmonic (induced) fields sum up coherently ($T_2 >> T_{plas}$), incoherently ($T_2<<T_{plas}$), or in an intermediate regime ($T_2\sim T_{plas}$) as schematically depicted in Figure \ref{fig4}. 

The decoherence effect is more important at shorter distances, at which the mutual interaction between LiCN and NP is stronger.
\noindent In intermediate conditions, i.e. $T_{2} = 50$ fs $\sim T_{plas}$, an increased excited-state population is found with respect to the fully coherent regime, since coherence, though still surviving, is partially suppressed by the presence of the surrounding environment.\\
\noindent Role of the polarization of the NP, induced by the external field $\vec{E}_{ext}$, is a fundamental ingredient in the plasmonic control of molecular absorption (Figure 4 in SI and related discussion), provided the magnitude of $\vec{E}_{ext}$ and $\vec{E}_{ind}$ (that depends on the detuning between molecular and plasmon excitation frequency, see Figure 3 in SI and related discussion) to be comparable. 

\begin{figure}[htbp]
\includegraphics[width=0.48\textwidth]{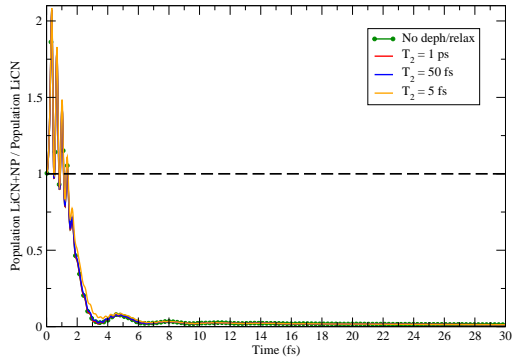}
\caption{Time evolution (up to 30 fs) of the ratio between the populations of LiCN+NP and bare LiCN with a Gaussian-envelope pulse (FWHM = 30 fs) and different dephasing regimes. \label{fig6}}
\end{figure}

It is clear from the analyses presented so far that the relative magnitude of the  electronic decoherence time  in the molecule $T_2$ vs the characteristic plasmon damping time $T_{plas}$ determines whether decoherence effects are relevant or not. Another fundamental time scale whose role must be clarified is the pulse duration. Are decoherence effects in the absorption enhancement still effective when a pulse of duration comparable to $T_2$ and $T_{plas}$, is applied to the molecule+NP system? We have therefore studied the possible interplay between the intrinsic time scale of the electronic dephasing (approximately $<$100 fs) and typical durations of the external pulse.
Indeed, coherence is generated in the system thanks to a combination of two effects, usually found in ultrafast spectroscopy experiments: fs-long pulse duration and use of a pulse frequency resonant with a molecular excitation.   \\
When  a tens-of-fs-long monochromatic pulse (Gaussian envelope with a FWHM of 30 fs) is used, decoherence is made ineffective in recovering absorption. This is shown in Figure \ref{fig6}, where we present the time evolution (up to 30 fs) of the ($\vert 2 \rangle$ + $\vert 3 \rangle$) normalized population for $T_2$ = 5 fs, 50 fs and 1 ps, and without dephasing. Calculations have been carried out for R=0.3 nm. Normalization refers to the population of the bare LiCN, in the same (de)coherence regime.
Using a monochromatic pulse with a Gaussian envelope (Eq. 11, FWHM = 30 fs) resonant with the molecular excitation under study, the effect of Gaussian width should be taken into account for a proper description of the molecular absorption, since it has the same order of magnitude of $T_{plas}$ and, possibly, $T_2$. 
Ratio between populations is always found less than unity, with the exception of the first 2 fs, which means that population of LiCN in presence of the NP is smaller than that of the bare LiCN. Absorption quenching is still due to the induced field on NP: comparison between the NP dipole obtained using the $\delta$ or the fs monochromatic pulse, with and without polarization induced by the external field, clearly shows the two pulses generate similar dipole amplitudes, responsible for the absorption quenching, according to the mechanisms explained above.  The result in Figure \ref{fig6} therefore corresponds to the suppressed absorption, already observed in the coherent regime (Eq. \ref{eq:coh}): the mechanism based on the incoherent sum of external- and induced-field contributions is contrasted by the coherence induced by the 30fs-pulse duration. 

\section{Conclusions}
\label{con}

We have investigated the role of electronic (de)coherence in controlling the molecular absorption, in presence of a metal NP. The specific geometric configuration employed here, with the transition dipole parallel to the NP surface and a broadband $\delta$ pulse interacting with the overall system, determines a strong quenching of the molecular absorption in coherent conditions, with a minimum at 1 nm.
Electronic decoherence plays a substantial role in modulating the  optical properties:
absorption is recovered (actually, even enhanced) at short distances (R $\le$ 1 nm), and the population minimum is shifted at a large distance:  a fast dephasing determines an incoherent sum  of external and induced field contributions, leading to an enhancement of the excited-state population by modifying the destructive interference between the molecular transition dipole moment and that induced in NP by the transition of interest.
Plasmonic modulation  of  molecular absorption is seen to be strongly modified when a fast dephasing induced by the external environment acts on the molecular state, thus possibly breaking the coherent interaction between external and induced fields. The external pulse duration has been seen to be essential to trigger the decoherence effect on the molecular absorption. At the same time, the detuning involving the molecular and plasmon excitations strongly affects the amplitude of the induced field on NP, thus playing a role in the time evolution of the molecular absorption.   

We have identified a number of conditions that are needed to magnify electronic decoherence effects in an ultrafast plasmonic spectroscopy:
\begin{itemize}
\item The relative effect of the external field and that induced on the molecule by the NP polarization must be of comparable strength. This may be realized playing with distances, molecular orientation and detuning.
\item Phenomena where in the coherent regime there is a subtraction of electric fields (i.e., a destructive interference) are better suited to reveal departure from such coherent regime, as in the incoherent counterpart such interference disappears. Phase induced radiative rate suppression has been already demonstrated experimentally.\cite{dul05}
\item The decay time of plasmons should be longer or at least comparable to the electronic decoherence rate of the chromophore. This limits the interesting metal nanostructures to those giving very sharp plasmonic peaks.
\item The pulse duration must be shorter than both $T_{plas}$ and $T_{2}$ (i.e., decay time of plasmon and decoherence time of molecules) otherwise the coherence is kept by the pulse itself.  
\end{itemize} 

These are the conditions that one has to realize to highlight the departure of a plasmonic experiment from the theoretical framework implicitly assuming coherence typically used to understand them. On the other hand, outside these conditions the coherent picture is preserved, which is important to know when ultrafast spectroscopies are combined with plasmonic nanostructure to achieve ultrasensitive time resolved spectroscopies.\cite{che13,wie14} \\

\begin{acknowledgments}
 Authors acknowledge funding from the ERC under the grant ERC-CoG-681285 TAME-Plasmons. Authors also acknowledge HPC Lab of University of Modena and Reggio Emilia and CINECA (Iscra C project "QNANO") for computational support.
\end{acknowledgments}

\section*{Supporting Information}

Ground-state (Figure 1) and excitation (Figure 2) energies of LiCN as a function of the NP distance. Results using a broadband $\delta$ pulse  (Figure 3). Effect of the induced field on LiCN populations (Figure 4). Role of detuning (Figures 5 and 6). 



\bibliography{bib}

\end{document}